\documentclass[%
superscriptaddress,
preprint,
showpacs,
amsmath,amssymb,
prc,
floatfix, ]%
{revtex4}
\usepackage{color}
\usepackage{graphicx}% Include figure files
\usepackage{dcolumn}% Align table columns on decimal point
\usepackage{bm}% bold math
\usepackage[dvipdfm,bookmarks=true,colorlinks,%
            citecolor=blue,linkcolor=blue,hypertex, %
            breaklinks=true]{hyperref}

\begin{document}

\title{Nuclear pairing reduction due to rotation and blocking}% Force line breaks with \\

\author{X. Wu }%
 \affiliation{State Key Lab of Nuclear Physics and Technology,
School of Physics, Peking University, Beijing 100871, China}
 \affiliation{School of Computing, University of Southern Mississippi,
Hattiesburg, MS, 39401, USA}
\author{Z. H. Zhang }%
 \affiliation{State Key Lab of Nuclear Physics and Technology,
School of Physics, Peking University, Beijing 100871, China}
 \affiliation{Institute of Theoretical Physics, Chinese Academy of Sciences,
Beijing 100190, China}
\author{J. Y. Zeng }%
 \affiliation{State Key Lab of Nuclear Physics and Technology,
School of Physics, Peking University, Beijing 100871, China}
\author{Y. A. Lei }
 \email{yalei@pku.edu.cn}
 \affiliation{State Key Lab of Nuclear Physics and Technology,
School of Physics, Peking University, Beijing 100871, China}

\date{\today}

\begin{abstract}
Nuclear pairing gaps of normally deformed and superdeformed nuclei are
investigated using the particle-number conserving (PNC) formalism
for the cranked shell model, in which the blocking effects are
treated exactly. Both rotational
frequency  $\omega$-dependence and seniority (number of unpaired particles)  $\nu$-dependence of
the pairing gap $\tilde{\Delta}$ are investigated. For the ground-state
bands of even-even nuclei, PNC calculations show that in general
$\tilde{\Delta}$ decreases with increasing $\omega$, but the
$\omega$-dependence is much weaker
than that calculated by the number-projected Hartree-Fock-Bogolyubov
approach. For the multiquasiparticle bands (seniority $\nu>
2$), the pairing gaps keep almost $\omega$-independent. As a
function of the seniority $\nu$, the bandhead pairing gaps
$\tilde{\Delta}(\nu,\omega=0)$ decrease slowly with increasing
$\nu$. Even for the highest seniority $\nu$ bands identified so far,
$\tilde{\Delta}(\nu,\omega=0)$ remains greater than $70\%$ of
$\tilde{\Delta}(\nu=0,\omega=0)$.
\end{abstract}

%21.60.-n 	Nuclear structure models and methods
%21.60.Cs 	Shell model
%23.20.Lv 	¦Ã transitions and level energies
%27.70.+q 	150 ¡Ü A ¡Ü 189
\maketitle

\section{Introduction}
\label{}

Since the seminal article by Bohr, Mottelson, and Pines~\cite{Bohr58},
significant effects of nuclear pairing were
established in fundamental nuclear properties~\cite{Bohr75}. Soon
afterwards, the Bardeen-Cooper-Schrieffer (BCS) theory for metallic
superconductivity and quasiparticle (qp) formalism were transplanted
in nuclear structure literature to treat nuclear pairing correlation~\cite{Migdal59, Belyaev59, Nilsson61}.
Now the BCS or more elaborate
Hartree-Fock-Bogolyubov (HFB) approximations are the standard
methods in nuclear physics. However, along with their great
successes, both BCS and HFB approximations for nuclear pairing
raise some concerns \cite{Zeng83, Molique97}. One of them
is the non-conservation of the particle-number. Because the number
of nucleons in a nucleus is not very large ($n\sim10^2$),
particularly the number of valence nucleons ($n\sim10$) dominating
the nuclear low-lying excited states is very limited, the relative
particle-number fluctuation, $\delta n/n$, is not negligible.
Indeed, it was found that in all self-consistent solutions to the
cranked HFB equation a pairing collapsing occurs for angular
momentum $I$ greater than a critical value $I_c$~\cite{Mottelson60}.

To restore this broken symmetry, many works have been done.
The Lipkin-Nogami (LN) method~\cite{Lipkin60, Lipkin61, Nogami64} has been quite extensively used in these years. After using this approximate particle-number projection method, pairing phase transition disappears~\cite{Gall94, Valor96}. However, earlier studies showed that the LN method broke down in the weak pairing limit~\cite{Dobaczewski93, Sheikh02}.
At the same time, various particle-number projection approaches of pairing interaction in BCS or HFB formalism have been developed~\cite{Canto85, Egido85, Anguiano01, Stoitsov07}.
In these approaches, the ideal treatment is variation after projection, but when spin goes higher, this method becomes very complicated and computational expensive. All these methods tried to solve the problem of the particle-number nonconservation. However, when it is achieved, no pairing phase transition was found~\cite{Canto85, Egido85}. This proves that the occurrence of nuclear pairing collapsing originates
from particle-number non-conservation. Other than the variational approach,
some methods directly solving the corresponding Sch\"{o}dinger equation have been developed~\cite{Zeng83, Pillet02}.
In these methods, the particle-number is strictly conserved. The particle-number conserving method used in Ref.~\cite{Pillet02}, in which the single-particle states stem from the Hartree-Fock mean field, is a little different from the method used in our work, in which the single-particle states stem from the Nilsson model.

Another problem related to the violation of particle-number conservation
is the occurrence of spurious states in the BCS (HFB)
qp formalism. As pointed out by Richardson~\cite{Richardson66},
an important class of low-lying excitations in
nuclei cannot be described in the standard BCS- or HFB-like
theories. The remedy in terms of the particle-number projection
considerably complicates the algorithm, yet failed to properly
describe the higher energy spectrum of the pairing Hamiltonian~\cite{Molique97}.

The most concerned issue is the proper treatment of the Pauli blocking
effect on pairing, which is responsible for the odd-even differences
in nuclear properties (binding energies, moments of inertia, etc.)
As emphasized by Rowe~\cite{Rowe70}, while the blocking effects are
straightforward, it is very difficult to treat consistently in the
qp formalism because they introduce different qp bases
for different blocked orbitals. Indeed, it was shown that the
properties of a rotational band are very sensitive to the Coriolis
response of the blocked single-particle orbitals~\cite{Zeng941}.

In this article, to investigate the pairing reduction due
to rotation and blocking we use the particle-number
conserving (PNC) formalism for treating the cranked shell model
(CSM) with pairing interaction, in which the particle-number is
conserved and the blocking effects are treated exactly. The details of the PNC formalism for calculating the MOI have been given in~\cite{Zeng942}. Only the
PNC formalism for calculating the nuclear pairing gap is given in
Sect.~II. Sect.~III gives PNC calculations for nuclear pairing gaps of
various types of pair-broken rotational bands in
normally deformed (ND) and superdeformed (SD) nuclei (seniority (number of unpaired particles) $\nu \geq 2$ for even-even nuclei, $\nu > 1$ for odd-$A$ nuclei), as well as the rotational frequency $\omega$-dependence and seniority
$\nu$-dependence of pairing gaps. A brief summary is given in Sect.~IV.

\section{PNC formalism for nuclear pairing gap}
The CSM Hamiltonian  of an axially deformed nucleus in the rotating frame is
\begin{eqnarray}
 H_\mathrm{CSM}
 & = &
 H_0 + H_\mathrm{P}
 = H_{\rm Nil}-\omega J_x + H_\mathrm{P}
 \ ,
 \label{eq:H_CSM}
\end{eqnarray}
where  $H_{\rm Nil}$ is the Nilsson Hamiltonian, $-\omega J_x$ is the
Coriolis interaction with cranking frequency $\omega$ about the $x$ axis,
$H_{\rm P} = H_{\rm P}(0) + H_{\rm P}(2)$ is the pairing
interaction,
\begin{eqnarray}
 H_{\rm P}(0)
 & = &
  -G_{0} \sum_{\xi\eta} a^+_{\xi} a^+_{\bar{\xi}}
                        a_{\bar{\eta}} a_{\eta}= -G_{0} \sum_{\xi\eta} s_\xi^+ s_\eta
  \ ,
 \\
 H_{\rm P}(2)
 & = &
  -G_{2} \sum_{\xi\eta} q_{2}(\xi)q_{2}(\eta)
                        a^+_{\xi} a^+_{\bar{\xi}}
                        a_{\bar{\eta}} a_{\eta}=-G_{2} \sum_{\xi\eta} q_{2}(\xi)q_{2}(\eta)s_\xi^+ s_\eta
  \ ,
\end{eqnarray}
where $\bar{\xi}$ ($\bar{\eta}$) labels the time-reversed state of a
Nilsson state $\xi$ ($\eta$), $q_{2}(\xi) = \sqrt{{16\pi}/{5}}
\langle \xi |r^{2}Y_{20} | \xi \rangle$ is the diagonal element of
the stretched quadrupole operator, and $G_0$ and $G_2$ are the
effective strengths of monopole and quadrupole pairing interactions respectively,
$s_\xi^+=a^+_{\xi} a^+_{\bar{\xi}}$ ($s_\eta= a_{\bar{\eta}} a_\eta$) is the pair creation (annihilation) operator.

In the PNC calculation, $H_{\rm CSM}$ is
diagonalized in a sufficiently large  cranked many-particle
configuration (CMPC) space \cite{Zeng942} and $| \Psi \rangle$ is expressed as
\begin{equation}
|\Psi\rangle=\sum_{i} C_i |i\rangle \qquad (C_i \; \textrm{real}), \label{eq:WF}
\end{equation}
where $|i\rangle$ is an eigenstate of $H_0$ with configuration
energy $E_i^{(0)}$, characterized by the particle-number $N$, parity
$\pi$, signature $r$ ($=e^{-i\pi\alpha}$) and seniority $\nu$ (number of unpaired particles). For the seniority $\nu=0$
ground state band ($K^\pi=0^+$) of an even-even
nucleus (qp vacuum in the BCS formalism), each $| i \rangle$ in Eq.~\ref{eq:WF} is of the
product form
\begin{equation}
s_\xi ^+ s_\eta ^+\cdots | 0 \rangle \ , \qquad \xi \neq \eta \neq \cdots \ .
\end{equation}
For the seniority $\nu=1$ band ($\sim$1-qp band in the BCS formalism) in an odd-even nucleus, $| i \rangle$ is of the form
\begin{equation}
a_\lambda ^+ s_\xi ^+ s_\eta ^+\cdots | 0 \rangle \ , \qquad \xi \neq \eta \neq \cdots (\neq \lambda) \ ,
\end{equation}
where $\lambda$ is the blocked single-particle state, $\xi \neq \eta \neq \cdots (\neq \lambda)$
(Pauli blocking effect) and the angular momentum projection along nuclear symmetry $z$-axis $K=\Omega_\lambda$.  For the seniority $\nu=2$ band ($\sim$2-qp band in the BCS formalism) in an even-even nucleus, $| i \rangle$ is of the form
\begin{equation}
a_\lambda ^+ a_\sigma ^+ s_\xi ^+ s_\eta ^+\cdots | 0 \rangle \ , \qquad \xi \neq \eta \neq \cdots (\neq \lambda \neq \sigma) \ ,
\end{equation}
where $\lambda \neq \sigma$ are two blocked single-particle states ($K=\Omega_\lambda+\Omega_\sigma$). The PNC forms of the $\nu > 2$ (multiquasiparticle) bands are similar. Strictly speaking, due to the Coriolis interaction $-\omega J_x$, $\nu$ and  $K$ are not exactly conserved for $\omega\neq0$. Walker and Dracoulis \cite{Walker99} pointed out that some forms of $K$-mixing must exist to enable the $K$-forbidden transition observed in a lot of low-lying rotational bands of axially symmetric nuclei. However, in the low-$\omega$ region, $\nu$ and $K$ may be served as useful quantum numbers characterizing a low-lying excited rotational band.

The kinematic and dynamic MOIs for the state $|\Psi\rangle$ are as follows~\cite{Zeng942, Zhang09}
\begin{equation}
J^{(1)}=\frac{1}{\omega}\langle\Psi|J_x|\Psi\rangle \ , \quad J^{(2)}=\frac{d}{d\omega}\langle\Psi|J_x|\Psi\rangle \ ,
\end{equation}
where
\begin{equation}
\langle\Psi|J_x|\Psi\rangle=\sum_{i}C_i^2\langle
i|J_x|i\rangle+2\sum_{i<j}C_iC_j\langle i|J_x|j\rangle \ ,
\end{equation}
is the angular momentum alignment of the state $|\Psi\rangle$.

In the PNC formalism, the nuclear pairing gap may be reasonably defined as~\cite{Canto85, Egido85, Shimizu89}
\begin{equation}
\tilde{\Delta}=G_0\left[-\frac{1}{G_0}\langle\Psi|H_{\rm
P}|\Psi\rangle\right]^{1/2} \ ,
\end{equation}
where $| \Psi \rangle$ is a PNC eigenstate (Eq.~\ref{eq:WF}) of $H_{\rm CSM}$ with eigenvalue $E$. In the BCS formalism for $H_{\rm CSM}$ with the monopole pairing interaction only, $H_{\rm P} = -G_0 S^+ S$, where $S^+=\sum_{\xi} s_\xi^+$, $S=\sum_{\eta} s_\eta$, and  for the qp vacuum band $| 0 \rangle \rangle$
\begin{equation}
|0\rangle\rangle=\Pi_\xi \left(U_\xi+V_\xi s_\xi^+\right)|0\rangle \ ,  \quad U_\xi^2 + V_\xi^2=1 \ ,
\end{equation}
$\tilde{\Delta}$ is reduced to the usual definition of nuclear pairing gap $\Delta$
\begin{equation}
\Delta = G_0 \langle\langle0|S^+|0\rangle\rangle= G_0 \sum_{\xi}U_\xi V_\xi \ .
\end{equation}

Calculations show that for the low-lying excited eigenstates of
$H_{\rm CSM}$, the number of important CMPC's (with weight $\geq
1\%$, say) is very limited (usually $<20$ for the ND
rare-earth nuclei), thus it is not difficult to get sufficiently
accurate solutions to the low-lying excited eigenstates of $H_{\rm
CSM}$  by diagonizing $H_{\rm CSM}$ in a sufficiently large CMPC
space~\cite{Zeng941, Zeng942}. To ensure the PNC calculations for nuclear low-lying excited
states both workable and accurate~\cite{Molique97, Wu89}, it is essential to adopt a CMPC truncation (Fock-space
truncation) in the PNC calculation in place of the usual
single-particle level (SPL) truncation in shell model calculations.
This is understandable from the perturbation expansion of
$H_{\rm CSM}$~(\ref{eq:H_CSM}), as it refers to a many-particle system with pairing
interactions. In general, the lower the configuration energy of the MPC is,
the larger the weight of the corresponding MPC in low-lying excited
eigenstates of $H_{\rm CSM}$ will be. The stability of the final results
with respect to the basis cut-off has been illustrated in details by
Molique and Dudek~\cite{Molique97}, as well as in~\cite{Liu02}.

In the following PNC calculations, $H_{\rm{CSM}}$ is diagonalized in
the CMPC space with dimension 1500 for both protons and neutrons.
The corresponding effective proton and neutron pairing strength
are adopted to reproduce the experimental odd-even
differences in nuclear binding energies. Proper Nilsson level
schemes are adopted to reproduce the experimental bandhead energies
and MOI of the low-lying excited seniority $\nu=1$ (1-qp)
bands. Thus, the pairing gaps $\tilde{\Delta}$ of
various low-lying excited bands can be convincingly extracted by
the PNC calculations without any free parameter.

\section{Calculations and discussions}
In this section the PNC calculations for nuclear pairing gaps of
some typical bands in ND and SD nuclei are presented. The
rotational frequency $\omega$-dependence and seniority
$\nu$-dependence of the pairing gaps are discussed in detail.

\subsection{Ground state bands of ${}^{168}$Yb and ${}^{168}$Hf}

The angular momentum dependence of pairing gaps  $\tilde{\Delta}_n$
(neutrons) and $\tilde{\Delta}_p$ (protons) for the ground state
band (gsb) of ${}^{168}$Yb and ${}^{168}$Hf have been calculated in
the number-projected HFB (NHFB) approach  in Ref.~\cite{Egido85}. The
pairing gap reductions in the observed angular momentum range
$I=0\rightarrow44\hbar$ ($\approx\omega=0.61$~MeV$/\hbar$) for
${}^{168}$Yb(gsb) and $I=0\rightarrow34\hbar$
($\approx\omega=0.52$~MeV$/\hbar$) for ${}^{168}$Hf(gsb) calculated
by NHFB are~\cite{Canto85, Egido85}
\begin{eqnarray}
\frac{\tilde{\Delta}_n(I=44
\hbar)}{\tilde{\Delta}_n(I=0)}&\approx&48\%, \quad
\frac{\tilde{\Delta}_p(I=44
\hbar)}{\tilde{\Delta}_p(I=0)}\approx63\%,  \quad \rm for\
 {}^{168}Yb(gsb)
\nonumber\\
\frac{\tilde{\Delta}_n(I=34
\hbar)}{\tilde{\Delta}_n(I=0)}&\approx&38\%, \ \quad \qquad \qquad \qquad \qquad \qquad\rm for\
{}^{168}Hf(gsb) \label{eq:DeltaYb}
\end{eqnarray}

\begin{figure}[h]
\centering
\includegraphics[scale=0.4]{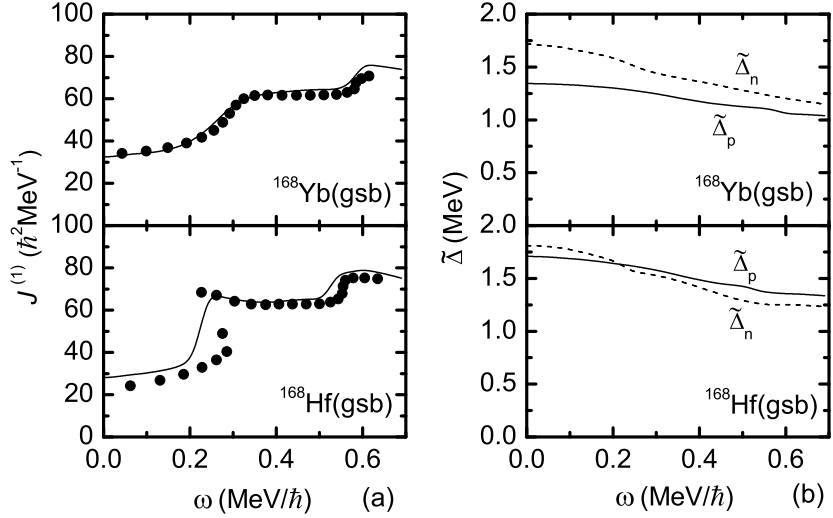}
\caption{\label{fig:YbHf} The MOIs and pairing gaps $\tilde{\Delta}$ for the ground
state bands of ${}^{168}$Yb and ${}^{168}$Hf. (a) The experimental
MOIs~\cite{Fitzpatrick95, Yadav08} are denoted by the solid circle $\bullet$, and the calculated
MOIs by the PNC method are denoted by solid lines. The Nilsson
parameters ($\kappa,\mu$) and deformation
($\varepsilon_2,\varepsilon_4$) are taken from~\cite{Nilsson69,
Bengtsson86}. The monopole and quadrupole pairing strengths for
protons and neutrons are adopted to reproduce the odd-even differences in nuclear binding energies, $G_n=0.30$,
$G_{2n}=0.010$, $G_p=0.29$ for ${}^{168}$Yb; $G_n=0.39$, $G_p=0.35$
for ${}^{168}$Hf. (b) The PNC calculated pairing gaps
for protons (neutrons) are denoted by solid
(dashed) lines.}
\end{figure}

For comparison, in this section the $\omega$-dependence of pairing
gaps of ${}^{168}$Yb(gsb) and ${}^{168}$Hf(gsb) are calculated using the PNC
formalism.
To validate the PNC calculations of  $\tilde{\Delta}_p$ (proton) and $\tilde{\Delta}_n$ (neutron) (Fig.~\ref{fig:YbHf}(b)), the kinematic MOIs $J^{(1)}$ are also calculated under the PNC formalism and compared with the experiments~\cite{Fitzpatrick95, Yadav08} (see Fig.~\ref{fig:YbHf}(a)). The experimental MOIs $J^{(1)}$ are very well reproduced by the PNC
calculations (except in the bandcrossing region). Thus, we believe
the PNC calculations of pairing gaps ($\omega$-dependence, $\nu$-dependence, etc.) are trustworthy. In the observed rotational frequency range, the pairing gap reductions calculated in PNC formalism are
\begin{eqnarray}
\frac{\tilde{\Delta}_n(\omega=0.61 \rm
MeV/\hbar)}{\tilde{\Delta}_n(\omega=0)}\approx70\%, \quad
\frac{\tilde{\Delta}_p(\omega=0.61 \rm
MeV/\hbar)}{\tilde{\Delta}_p(\omega=0)}\approx80\%, \quad \rm for \
{}^{168}Yb
\nonumber\\
 \frac{\tilde{\Delta}_n(\omega=0.52
\rm MeV/\hbar)}{\tilde{\Delta}_n(\omega=0)}\approx70\%, \quad
\frac{\tilde{\Delta}_p(\omega=0.52 \rm
MeV/\hbar)}{\tilde{\Delta}_p(\omega=0)}\approx83\%, \quad \rm for \
{}^{168}Hf
\end{eqnarray}
which remains more than $70\%$ of the bandhead value in all
experimental $\omega$ range. As expected,
in both NHFB and PNC formalism no pairing phase transition from
superfluidity to normal motion ($\tilde{\Delta}\rightarrow0$) is
found with increasing $\omega$. However, the $\omega$-dependence of
$\tilde{\Delta}$ in PNC calculations is weaker than that calculated by NHFB approach. By the way, it is noted that due to the neutron sub-shell effect at $N=98$, in both
PNC and NHFB calculations for ${}^{168}$Yb, the pairing gap
reduction of neutron is larger than that of proton.
It was noted by I. Hamamoto~\cite{Hamamoto76} that an inherent issue of CSM is the violation of rotational symmetry, and the reliability of calculations in the CSM, particularly in the bandcrossing region, is questionable. Afterwards, the angular momentum projection techniques were developed~\cite{Hara95}. It is interesting to note that the $\omega$-dependence of the pairing gaps for the gsb of  ${}^{168}$Yb calculated by the angular momentum projection technique~\cite{Sun94} are similar to that of the PNC calculations.

\subsection{Multiquasiparticle bands of the heavier rare-earth nuclei ($A\sim178$)}
\begin{figure}[h]
\centering
\includegraphics[scale=0.4]{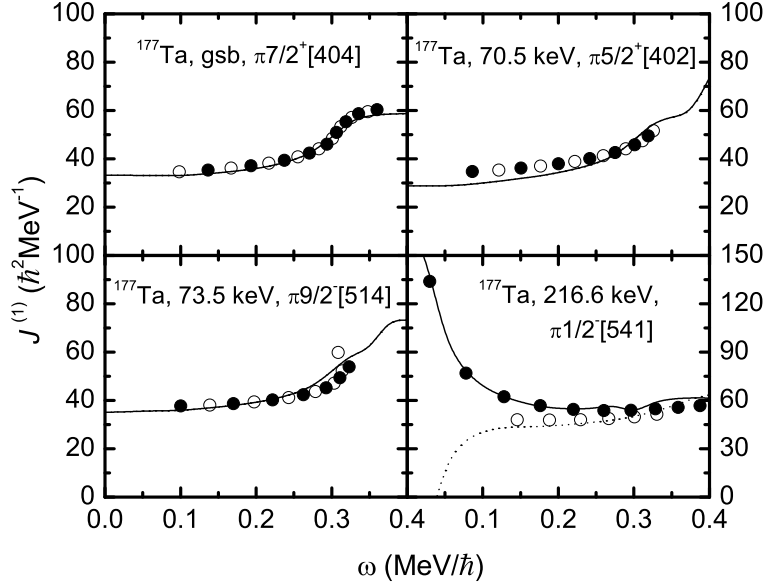}
\caption{\label{fig:Ta} MOIs of four low-lying  seniority $\nu=1$ bands in ${}^{177}$Ta. The
experimental MOIs~\cite{Dasgupta00} are denoted by $\bullet$
$(\alpha=1/2)$ and $\circ$ $(\alpha=-1/2)$, respectively. The
calculated MOIs by the PNC method are denoted by solid lines
($\alpha=1/2$) and dotted lines ($\alpha=-1/2$), respectively. The
Nilsson parameters ($\kappa,\mu$) in~\cite{Nilsson69} are slightly
adjusted to reproduce the bandhead energies of the 1-qp bands. For
protons, $\kappa_4=0.060$ ($N=4$), $\kappa_5=0.061$ ($N=5$),
$\mu_4=0.55$, $\mu_5=0.69$. For neutrons, $\kappa_5=0.066$,
$\kappa_6=0.058$, $\mu_5=0.49$, $\mu_6=0.40$. The deformation
parameters ($\varepsilon_2,\varepsilon_4$)=(0.24, 0.04) are from~\cite{Bengtsson86}, i.e., an average of the neighboring
even-even Hf and W isotopes. The effective pairing interaction
strengths for both proton and neutron, $G_n=0.26$~MeV, $G_p=0.26$~MeV, are
determined by the experimental odd-even differences in nuclear binding energies.}
\end{figure}

The seniority $\nu$-dependence of nuclear pairing gaps have been
investigated by Dracoulis et al., using the LN method~\cite{Dracoulis98}. They showed that
the bandhead pairing gap $\Delta(\nu,\omega=0)$ decreases
approximately by
\begin{equation}
\Delta(\nu, \omega=0)=(0.75)^{\nu/2}\Delta(\nu=0, \omega=0). \label{eq:LNDelta}
\end{equation}
In this section we will investigate the $\nu$-dependence of the
$\tilde{\Delta}$ using the PNC formalism. To get pairing gaps for
these multiquasiparticle bands, a proper Nilsson level scheme for
the deformed heavier rare-earth nuclei ($A\sim178$) is necessary.
However, the level scheme (Lund systematics)~\cite{Nilsson69, Bengtsson86}
is unable to properly reproduce the
experimental bandhead energies of the low-lying excited 1-qp bands
of ${}^{177}$Ta, particularly the gsb, $\pi7/2^{+}[404]$. So the Nilsson
parameters $(\kappa, \mu)$ in~\cite{Nilsson69} are slightly adjusted
(see the caption of Fig.~\ref{fig:Ta}).
Fig.~\ref{fig:Ta} shows the experimental~\cite{Dasgupta00} and calculated MOIs of four 1-qp bands in ${}^{177}$Ta.

\begin{figure}[h]
\centering
\includegraphics[scale=0.25]{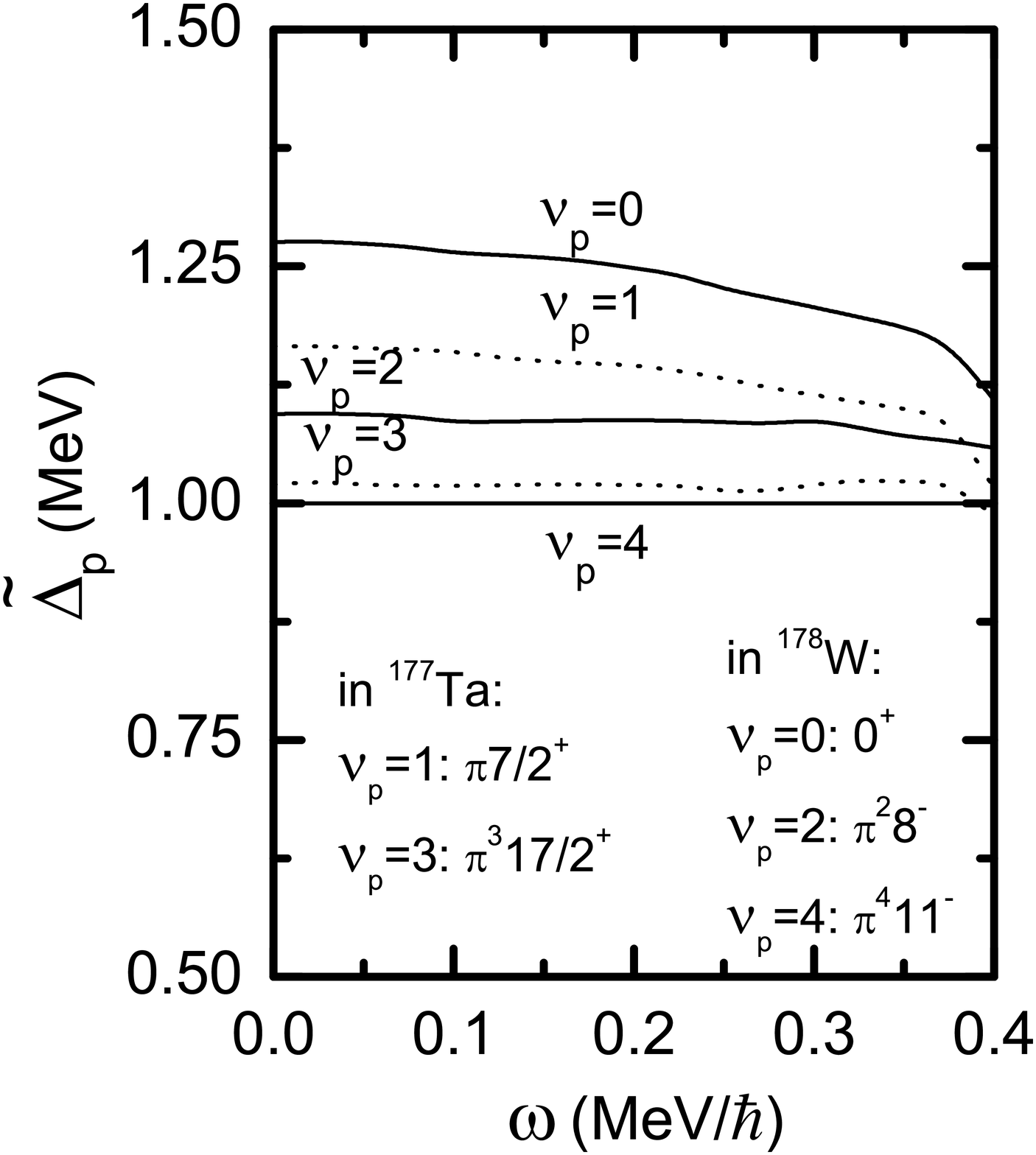}
\caption{\label{fig:TaDelta}The protons pairing gaps $\tilde{\Delta}_p$
for the $\nu_p=1, 3$ bands in ${}^{177}$Ta (dotted lines) and $\nu_p=0, 2, 4$ configurations in ${}^{178}$W (solid lines).${}^{177}$Ta: $\nu_p=1$ band (gsb), $\pi7/2^+[404]$; $\nu_p=3$, $K^\pi=17/2^+$
band at 1523 keV~\cite{Dasgupta00},
$\pi^317/2^+(7/2^+[404]\otimes9/2^-[514]\otimes1/2^-[541])$.
${}^{178}$W:  gsb, $\nu_p=0$, $K^\pi=0^+$; $\nu_p=2$ configuration
$\pi^28^-(7/2^+[404]\otimes9/2^-[514])$ in $K^\pi=15^+$
($\pi^28^-\otimes\nu^27^-$) band at 3653 keV; $\nu_p=4$ configuration
$\pi^411^-(7/2^+[404]\otimes5/2^+[402]\otimes9/2^-[514]\otimes1/2^-[541])$
in $K^\pi=18^+$ ($\pi^411^-\otimes\nu^27^-$) band at 4878 keV
\cite{Purry98, Dracoulis98}.
}
\end{figure}

In Fig.~\ref{fig:TaDelta}, we show the PNC calculations of the proton
pairing gaps for $\nu=1, 3$ bands in ${}^{177}$Ta and the gsb, and  multiquasiparticle bands
with $\nu_p=2$ and 4 proton configurations in ${}^{178}$W~\cite{Dasgupta00,
Purry98, Dracoulis98}. The $\omega$-dependence of $\tilde{\Delta}_p$
for ${}^{178}$W(gsb) is similar to that of ${}^{168}$Yb(gsb) and
${}^{168}$Hf(gsb) (see Fig.~\ref{fig:YbHf}(b)). The $\omega$-dependence of
$\tilde{\Delta}_p$'s of the four low-lying excited 1-quasiproton bands
of ${}^{177}$Ta is similar to ${}^{178}$W(gsb), so
only the $\tilde{\Delta}_p$ for the gsb ($\pi7/2^+[404]$) of
${}^{177}$Ta is shown in Fig.~\ref{fig:TaDelta}. The $\omega$-dependence of the
$\nu_p=2$ configuration $\pi^28^-$ ($7/2^+[404] \otimes 9/2^-[514]$) in ${}^{178}$W is
weaker. For the low-lying excited multiquasiparticle (seniority
$\nu_p > 2$) bands, $\tilde{\Delta}_p$'s keep \emph{nearly
$\omega$-independent}. In fact, for realistic nuclei, the blocking
effects on pairing are significant only for a few orbitals nearest
to the Fermi surface. For low-lying excited multiquasiparticle
bands, a few orbitals nearest to the Fermi surface are almost
blocked, and for orbitals far from the Fermi surface the
$\omega$-dependence of the blocking effects on pairing are quite
small.

As a function of seniority $\nu$, in general, the pairing gap
$\tilde{\Delta}(\nu)$ gradually decreases with increasing $\nu$. The
pairing gap reductions at bandhead ($\omega=0$) calculated by PNC
method are
\begin{eqnarray}
\frac{\tilde{\Delta}_p(\nu=1)}{\tilde{\Delta}_p(\nu=0)}\approx91\% \ ,\quad
\frac{\tilde{\Delta}_p(\nu=2)}{\tilde{\Delta}_p(\nu=0)}\approx86\% \ , \nonumber\\
\frac{\tilde{\Delta}_p(\nu=3)}{\tilde{\Delta}_p(\nu=0)}\approx80\% \ , \quad
\frac{\tilde{\Delta}_p(\nu=4)}{\tilde{\Delta}_p(\nu=0)}\approx78\% \ . \label{eq:DeltaTa}
\end{eqnarray}
which are weaker than that given in Eq.~(\ref{eq:LNDelta})~\cite{Dracoulis98}.
Even for the highest seniority $\nu$ bands identified so far, the
pairing gaps $\tilde{\Delta}(\nu,\omega=0)$ is always larger than $70\%$
of $\tilde{\Delta}(\nu=0,\omega=0)$.

\subsection{Ground state bands of ${}^{238}$U and ${}^{253}$No}
\begin{figure}[h]
\centering
\includegraphics[scale=0.4]{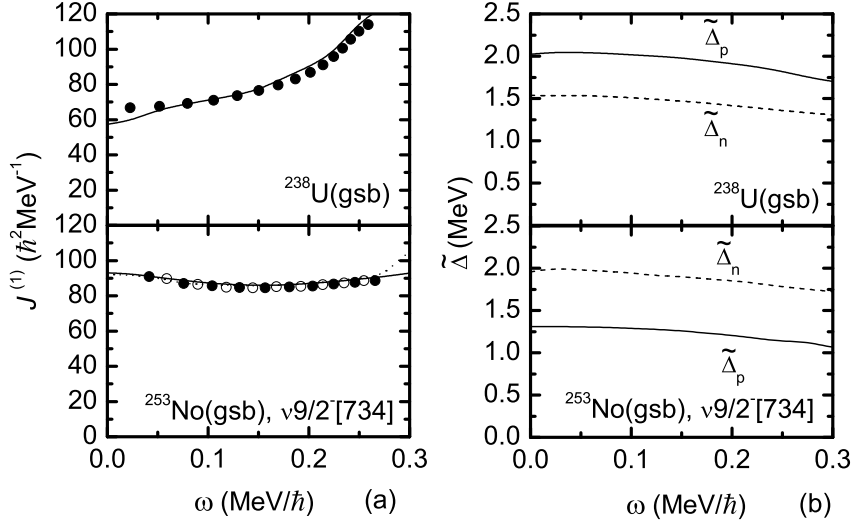}
\caption{\label{fig:UNo}The MOIs and pairing gaps of the gsb of
${}^{238}$U and ${}^{253}$No. (a) The
experimental MOIs~\cite{Firestone98, Reiter05} are denoted by solid circle $\bullet$ ($\alpha=0,
1/2$) and open circle $\circ$ ($\alpha=-1/2$). The calculated MOIs
by the PNC method are denoted by solid lines ($\alpha=0, 1/2$) and
dotted lines ($\alpha=-1/2$). The Nilsson parameters ($\kappa, \mu$)
are taken from \cite{Nilsson69} both for ${}^{238}$U and
${}^{253}$No. Deformations ($\varepsilon_2, \varepsilon_4)=(0.21, -0.04)$ of
${}^{238}$U are taken from~\cite{Bengtsson86}. For ${}^{253}$No, the quadrupole
deformation parameters for the neighbor even-even nuclei are deduced
as $\beta_2=0.28\pm0.02$ from the experiment~\cite{Herzberg01}. Here
we choose $\varepsilon_2=0.26$ and and $\varepsilon_4=0.01$ in our
calculation for ${}^{253}$No. To reproduce the experimental odd-even
difference in binding energies and the $\omega$-dependence of MOIs,
it seems that a small amount of quadrupole pairing interaction is
necessary, i.e., $G_n=0.29$~MeV, $G_{2n}=0.015$~MeV, $G_p=0.32$~MeV,
$G_{2p}=0.040$~MeV for ${}^{238}$U; $G_n=0.22$~MeV, $G_{2n}=0.010$~MeV,
$G_p=0.26$~MeV, $G_{2p}=0.010$~MeV for ${}^{253}$No. (b) The PNC calculated pairing gaps
$\tilde{\Delta}_p$'s ($\tilde{\Delta}_n$'s) for the gsb's of ${}^{238}$U and ${}^{253}$No are denoted by solid (dashed) lines.}
\end{figure}

The PNC calculations for the gsb's of actinide nuclei, ${}^{238}$U
and ${}^{253}$No, are shown in Fig.~\ref{fig:UNo}. The experimental data of MOIs
are taken from~\cite{Firestone98, Reiter05}. The PNC calculations
for the pairing gap reduction show
\begin{eqnarray}
&&\frac{\tilde{\Delta}_n(\omega=0.30 \rm
MeV/\hbar)}{\tilde{\Delta}_n(\omega=0)}\approx85\% \ , \quad
\frac{\tilde{\Delta}_p(\omega=0.30 \rm
MeV/\hbar)}{\tilde{\Delta}_p(\omega=0)}\approx84\% \ , \quad \rm for \
{}^{238}U(gsb)
\nonumber\\
&&\frac{\tilde{\Delta}_n(\omega=0.30 \rm
MeV/\hbar)}{\tilde{\Delta}_n(\omega=0)}\approx84\% \ , \quad
\frac{\tilde{\Delta}_p(\omega=0.30 \rm
MeV/\hbar)}{\tilde{\Delta}_p(\omega=0)}\approx76\% \ , \quad \rm for \
{}^{253}No(gsb)
\end{eqnarray}
i.e., $\tilde{\Delta}_p$'s and $\tilde{\Delta}_n$'s decrease very
slowly with increasing $\omega$, quite similar to the rare-earth
nuclei.

\subsection{Pairing gaps of SD bands}
\begin{figure}[h]
\centering
\includegraphics[scale=0.4]{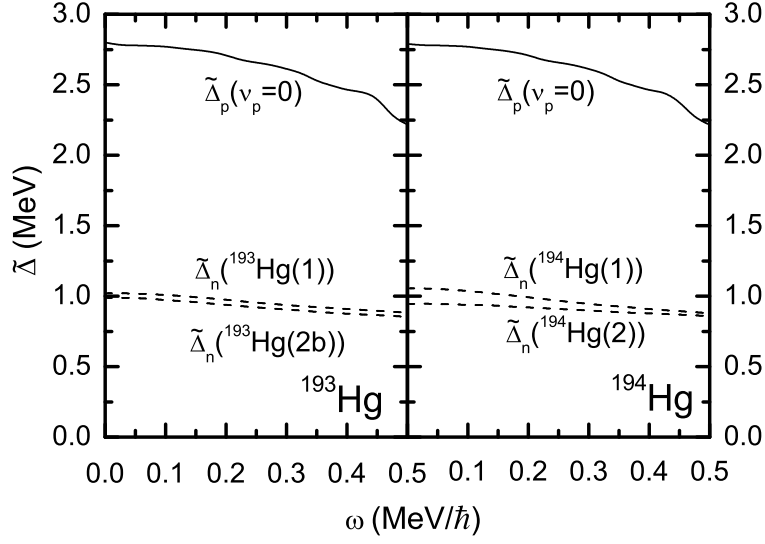}
\caption{\label{fig:HgDelta}The pairing gaps of SD bands in Hg isotopes calculated by
the PNC formalism. The Nilsson level scheme are taken from~\cite{Bengtsson85}.
${}^{193}$Hg(1): 1-quasineutron SD band,
$\nu5/2^-[512], \alpha=-1/2$; ${}^{193}$Hg(2b): 1-quasineutron SD
band, $\nu9/2^+[624], \alpha=1/2$. ${}^{194}$Hg(1): quasi-vacuum SD
band ($\alpha=0$); ${}^{194}$Hg(2): 2-quasineutron SD band
$\nu^27^-(5/2^-[512]\otimes9/2^+[624]), \alpha=0$.}
\end{figure}

Chasman~\cite{Chasman90} pointed out that BCS treatment of nuclear
pairing is not appropriate for SD states because the single-particle
level density near the Fermi surface is low and the BCS method is
not correct in this limit. In most cases, the linkage between the SD bands and low-lying excited states have not yet been established, thus the actual spins of some SD bands are not determined, but the dynamic MOIs $J^{(2)}$ can be extracted from the observed differences in subsequent $\gamma$ transition energies,
\begin{equation}
J^{(2)} (I) = 4\hbar^2 / \left[ E_\gamma (I+2\rightarrow I)-  E_\gamma (I\rightarrow I-2) \right] \ .
\end{equation}
Obviously, the accuracy of $J^{(2)}$ is lower than $J^{(1)}(I) = (2I+1) \hbar^2 / \left[ E_\gamma (I+1\rightarrow I-1) \right]$. However, the actual spins of some SD bands have been established experimentally (e.g., see Ref.~\cite{Bouneau98}), thus the  $J^{(1)}$'s can be accurately extracted.
The $\omega$-dependence of experimental
MOIs for series of SD bands~\cite{Riley90, Joyce94, Han99} were
reproduced very well by the PNC calculations for the CSM with both
monopole and quadrupole pairing interactions~\cite{Liu02, Xin00,
Liu04}. In Fig.~\ref{fig:HgDelta}, the PNC calculations of the pairing gaps
$\tilde{\Delta}$ for the $\nu_n=1$ SD bands ${}^{193}$Hg(1)
($\nu5/2^-[512], \alpha=-1/2$), ${}^{193}$Hg(2b) ($\nu9/2^+[624],
\alpha=1/2$), the $\nu=0$ SD band ${}^{194}$Hg(1), and
$\nu_n=2$ SD band ${}^{194}$Hg(2) ($\nu^27^-,
5/2^-[512]\otimes9/2^+[624], \alpha=0$) are shown in Fig.~\ref{fig:HgDelta}. The
Nilsson level scheme are taken from~\cite{Bengtsson85}. PNC
calculations show that:

(a) For SD bands in Hg isotopes,
$\tilde{\Delta}_p$(proton)$\gg\tilde{\Delta}_n$(neutron), which is
caused by the large gap at $Z=80$ in the proton Nilsson level scheme of
SD Hg isotopes.

(b) For SD bands, no pairing collapsing is found with
increasing $\omega$ either. For ${}^{194}$Hg(1) ($\nu=0$, SD band),
the pairing gap reduction with increasing $\omega$ is,
\begin{equation}
\frac{\tilde{\Delta}_n(\omega=0.50 \rm
MeV/\hbar)}{\tilde{\Delta}_n(\omega=0)}\approx83\% \ , \
\frac{\tilde{\Delta}_p(\omega=0.50 \rm
MeV/\hbar)}{\tilde{\Delta}_p(\omega=0)}\approx80\% \ .
\end{equation}
For both the $\nu_n=1$, SD bands ${}^{193}$Hg(1) and
${}^{193}$Hg(2b)
\begin{equation}
\frac{\tilde{\Delta}_n(\omega=0.50 \rm
MeV/\hbar)}{\tilde{\Delta}_n(\omega=0)}\approx86\% \ .
\end{equation}
For the
$\nu_n=2$, SD band ${}^{194}$Hg(2)
\begin{equation}
\frac{\tilde{\Delta}_n(\omega=0.50 \rm
MeV/\hbar)}{\tilde{\Delta}_n(\omega=0)}\approx90\% \ .
\end{equation}

\section{Summary}
The $\omega$- and $\nu$-dependence of the nuclear pairing gaps of
multiquasiparticle bands in well-deformed and SD nuclei are
calculated under the PNC formalism, in which the blocking effects on pairing are exactly taken into
account. PNC calculations show that the $\omega$-dependence of
pairing gaps $\tilde{\Delta}$ for the $\nu=0$ (qp-vacuum) bands is weaker than
what predicted in the particle-number projected HFB formalism.
For the low-lying excited $\nu>2$ ($\sim$multiquasiparticle) bands, $\tilde{\Delta}_p$'s and $\tilde{\Delta}_n$'s keep
almost $\omega$-independent. As a function of seniority $\nu$, the
bandhead pairing gaps $\tilde{\Delta}(\omega=0,\nu)$, decrease
slowly with increasing $\nu$. Even for the highest seniority bands
identified so far, the pairing gaps $\tilde{\Delta}_p(\omega=0,
\nu)$ and $\tilde{\Delta}_n(\omega=0,\nu)$ remains larger than $70\%$
of the bandhead value of the qp-vacuum band.

\section{Acknowledgement}
This work is supported by the Natural Science Foundation of China
under the Nos.~10976001, 10935001, and the 973 program
2008CB717803.

%\bibliography{reference}

\end{document}